\documentstyle[aps,prl,preprint,balanced,epsfig]{revtex}
\begin{document}
\title{Large Scale-Invariant Fluctuations in Normal Blood Cell Counts: A sign of criticality? }
\author{Carlos A. Perazzo, Elmer A. Fernandez, Dante R. Chialvo$^{\dagger}$, and Peter Willshaw}
\address{{\sl Universidad Favaloro. Sol\'\i s 453, (1078) Buenos Aires - Argentina, 
$^{\dagger}$Center for Studies in Physics and Biology,
The Rockefeller University, 1230 York Avenue, New York NY10021, USA}}
\date{\today}
\maketitle

\begin{abstract} 
All types of blood cells are formed by differentiation from a small self-maintaining population 
of pluri-potential stem cells in the bone marrow. Despite abundant information on the 
molecular aspects of division, differentiation, commitment and maturation of these cells, 
comparatively little is known about the dynamics of the system as a whole, and how it works 
to maintain this complex ``ecology'' in the observed normal ranges throughout life. Here we 
report unexpected large, scale-free, fluctuations detected from the first long-term analysis of 
the day-to-day variability of a healthy animal's blood cell counts measured over one thousand 
days. This scale-invariance cannot be accounted for by current theoretical models, and 
resembles some of the scenarios described for self-organized criticality.
\end{abstract}

\vspace{.1truein} 

{\bf Running Title: Scale-Invariance in blood cells counts} \\

Address all correspondence to:\\
Dr. Peter Willshaw, Universidad Favaloro. Sol\'\i s 453, (1078) Buenos Aires - Argentina\\
Email: peterw@favaloro.edu.ar Phone: +54-11-43849920 Fax: +54-11-43839732\\ 
\newpage


The existence of scale-invariant fluctuations in nature is well-documented\cite{bunde}. Scale-invariance 
refers \cite{bunde,feder} to the absence of a single time or length scale characteristic of the (temporal or spatial) 
pattern of change. In such cases fluctuations happen at all scales, unlike the more familiar cases 
in which there is a simple statistical spread around a defined mean value. Scale-invariance has 
been shown for fluctuations in heart rate\cite{meesmann}, gait dynamics\cite{hausdorff}, DNA nucleotide sequences
\cite{peng1}, and lung inflation\cite{suki}. 
How these fluctuations are generated by regulatory physiological mechanisms and 
what their relevance is in biology is the focus of much current interest. In addition, 
it has been argued that 
scale-invariance in itself can be the fingerprint of a biological system poised at a critical state, 
resulting in  improved dynamics with robust tolerance to error, and at the same time remaining very susceptible
 to change\cite{bak}.  Roughly 
speaking, this idea of self-organized criticality as a homeostatic mechanism is that systems 
with many interacting non-linear parts could find themselves moving toward the critical state 
(i.e. ``self-organizing'') simply because it would be absolutely impractical to administer it in 
any centralized way.
Here we show that the regulation of the number of the formed elements in blood also 
has scale-invariant properties. We found that the day-to-day variability in the number of 
platelets (PLA), red (RBC) and white blood cells (WBC) from two healthy sheep observed for one 
thousand days is scale-invariant over a range spanning from a few to over two hundred days.

Even in healthy subjects\cite{mackey,dale} the number of blood cells fluctuates from one day to the next. Health 
practitioners are very aware of this fact, to the point that these quantities are usually described as ranges 
or maximal limits and not as means and standard deviations. In other words, for these variables 
normality could be defined in one sense as ``fluctuations within a normal range''.
Figure 1 shows the daily values (over 1000 days) for blood cells taken from two healthy sheep. Overall, 
the values observed are consistent with the normal range reported\cite{jain} in the literature for these animals. 
Simple inspection of the data reveals fast, medium and slow fluctuations, and a certain degree of 
correlation in the behaviour of the time series between the animals, as if they were responding in a 
similar way to some common external factor.
What is the nature of the extensive fluctuations seen in all the series plotted in Figure 1? 
According to current understanding\cite{mackey}, homeostatic mechanisms act to compensate the loss of a given 
number of cells by speeding up the process of maturation of precursors, i.e., there is a negative 
feedback loop attempting to maintain constant the number of blood cells. The precise characterization 
of the observed fluctuations is important in assessing the adequacy of current models, because much is 
known about the dynamics which regulatory mechanisms of this kind are capable of exhibiting\cite{mackey}. Thus 
one could reject or accept classes of models, or propose the need for novel ones, depending on the type 
of fluctuations found experimentally.

The dynamics of any underlying control process is expected to be reflected in the temporal 
correlation between the data points; this is the physiologically important information we are interested 
in. We consider here two extreme (null) hypotheses: the first is to assume a tight feedback mechanism 
perturbed by some jitter where one should simply see a mean value contaminated with white noise. The 
second corresponds to the very unlikely case that no control is exerted to compensate for blood cell 
losses, resulting in a random walk motion, in which the fluctuation is just generated by the day-by-day 
summation of statistically independent increments or decrements.
To compare our observed time series against these two possibilities, surrogate data sets were 
constructed to mimic the null hypotheses cited above.

In Figure 2, the raw time series of red blood cells 
of one animal is plotted in the top panel. Panels denoted S1 and S2 depict the surrogate series of the 
first and second types respectively. The first one (S1) is constructed by re-ordering at random the raw 
values of the top panel, like shuffling a deck of cards. In this way, any temporal relationship between 
the counts on one day and the next in the original data is broken, and the resulting time series is white 
noise. Surrogate S2 is built by (1) generating an accessory file with the day-to-day differences of the 
raw data in the top panel; (2) randomizing the order of the data points from this accessory file, and (3) 
integrating the result over the whole length of the file to synthesize a new time series. Surrogate type 2 
is therefore a random walk-like signal. By using different random orderings, one can build many 
examples or realizations of these types of surrogates.

To quantify the characteristics of the temporal correlations in the data we use detrended 
fluctuation analysis (DFA)\cite{peng2}. This method has been used successfully to analyze biological time series, 
circumventing some technical difficulties, such as trends and non-stationarities. The DFA algorithm 
involves the following steps: 1) Denoting $C_{j}$ as the blood cell daily count of interest on the $j_{-th}$ day, 
produce a new time-integrated series $y_{i}=\sum_{j=1}^i (C_{j} -C_{av})$, where $C_{av}$ is 
the average of all $C_{j}$. 2) The new 
series $y_{i}$ is subsequently divided into boxes of equal length $n$, and in each box a least-squares line is 
fitted to the data, representing the trend over the chosen time interval. 3) Define and calculate $F(n)$ as 
the square root of the mean of the squares of the residuals in all boxes of length n for a given value of 
n. Thus, $F(n)$ quantifies the magnitude of the fluctuation of the integrated and detrended time series 
over increasingly long time intervals. The presence of a straight line of slope $\alpha$ in the log-log 
plot of  F(n) vs. n implies a relationship of the type $F(n)\propto n^{\alpha}$ . 
The range of $n$ over which this relationship 
holds defines the time span over which the fluctuations are scale-invariant. The two extreme cases 
considered have expected values of $\alpha = 0.5$ for white noise, and 1.5 for a random-walk motion.

Panel a in Fig. 3 is a log-log plot of F(n) vs. n resulting from the analysis of the red cells 
data set plotted in Fig. 2. Symbols on the right of the figure indicate to which time series the calculated 
function belongs, R denoting the raw time series, and S1 and S2 the 30 random realizations of 
surrogates 1 and 2 respectively. The plot for the raw data is approximately linear over time lags n from 
1-200 days, indicating that blood cell count fluctuations are scale-invariant. We found similar scaling 
behaviour in all series in both sheep. The lines of best fit gave slopes $\alpha= 0.98$ and 1.00 for RBC, 
$\alpha = 1.24$ and 1.11 for PLA, and $\alpha = 0.83$ and 0.83 for WBC (sheep 77 and 78 respectively). The lowest 
value of the correlation coefficient r for all raw data series was 0.995. Panel b in Figure 3 illustrates an 
important consequence of scale-invariance: on increasing the observation time by a factor of $k$ the 
fluctuations are found to be $k^{\alpha}$ larger, that is to say the longer we look the bigger 
the range of values we will find.

Additional insight can be gained by applying standard spectral analysis techniques to the same 
series. In this case, it is appropriate to consider the spectral analysis of the daily differences 
(i.e, $C_{j} - C_{j-1} )$ of the time series, since this is of a stationary nature by construction.
 We computed the power spectra $S(f)$ of each differentiated time series and in all cases we found that the square of their amplitudes 
scales as $S(f) \sim f^{\beta}$ . The scaling factors $\alpha$ and $\beta$ are related\cite{bunde} 
as $\beta =3- 2\alpha$.
 The values obtained via spectral analysis (for sheep 77 and 78) were $\beta= 0.97$ and 0.91 for RBC, 
(the values predicted from $\alpha$ are 1.04 and 1.00 respectively); for PLA $\beta = 0.65$ and 0.70 
(predicted values of 0.52 and 0.78 respectively); and for WBC $\beta = 1.14$ and 1.16 
(1.34 and 1.34 respectively). There is clear agreement between the scaling exponents obtained using both methods.

The value of $\beta$ provides additional information regarding the nature of the process, because 
 $0 < \beta < 1$   implies that a high value is more likely to be follow by a small value and viceversa,
 i.e., there is anti-correlation between values collected on successive days. The fact that we found a wide range (1-
200 days) over which this scaling factor holds, indicate that the mechanism responsible for the anti-
correlated dynamics operates at all time scales, something that a regulatory system as a
simple negative feedback-like loop is incapable of doing.
 
We must therefore reject the two null hypotheses considered. The fact that the data is not white is 
surprising, since this is the most likely possibility if there is simple feedback control such as that 
postulated to relate erythropoietin production and erythrocyte number. The fact that the data is not a 
random walk does at least suggest that control is present, and that it is non-trivial.
 Substantial work exists regarding the architecture of haematopoietic regulation, including the 
seminal work of Mackey\cite{mackey} but interestingly enough no model has as yet reproduced the scale-invariance 
reported here. We suspect that an important part of the fluctuations we have observed reflects genuine 
intrinsic dynamics of the system, which possesses a very large number of parts. To account for these 
observations, models which preserve this very large number of degrees of freedom are needed.
The behaviour here reported for sheep has also been suggested\cite{ary} to be present in human 
neutrophil counts, albeit using a far shorter time series than ours. To show that this type of scale 
invariance is indeed present in humans without going to the extreme of daily sampling, it could
be possible to 
collect data from many individuals who have been subject to blood sampling and 
counting on at least two occasions a known number of days, months or years apart and estimating a 
similar fluctuation function. This could be achieved for example in records of regular blood donors, 
and will estimate the likelihood of encountering a given difference as a function of time lag. At present 
no information exists which health practitioners can use to forecast such simple physiological 
parameters in situations of interest such as bone marrow transplant.

 If our results hold in humans, one important consequence of scale invariance is to make it 
difficult to differentiate between the response to a given treatment and a chance value due to the 
intrinsic (scale-invariant) nature of this type of fluctuations, especially if the patient is seen at long 
intervals. Health practitioners generally make clinical decisions on the basis of isolated data points, it 
being implicit that waiting longer or taking more samples can only reduce the variance. Our findings 
imply exactly the opposite; waiting longer increases the variance!

Acknowledgements: We are indebted to  veterinarians Drs. M. Besancon, P. Iguain, and M. Tealdo for
the collection and  processing of the blood samples.

\newpage

\begin{figure}[htbp]
\centerline{\psfig{figure=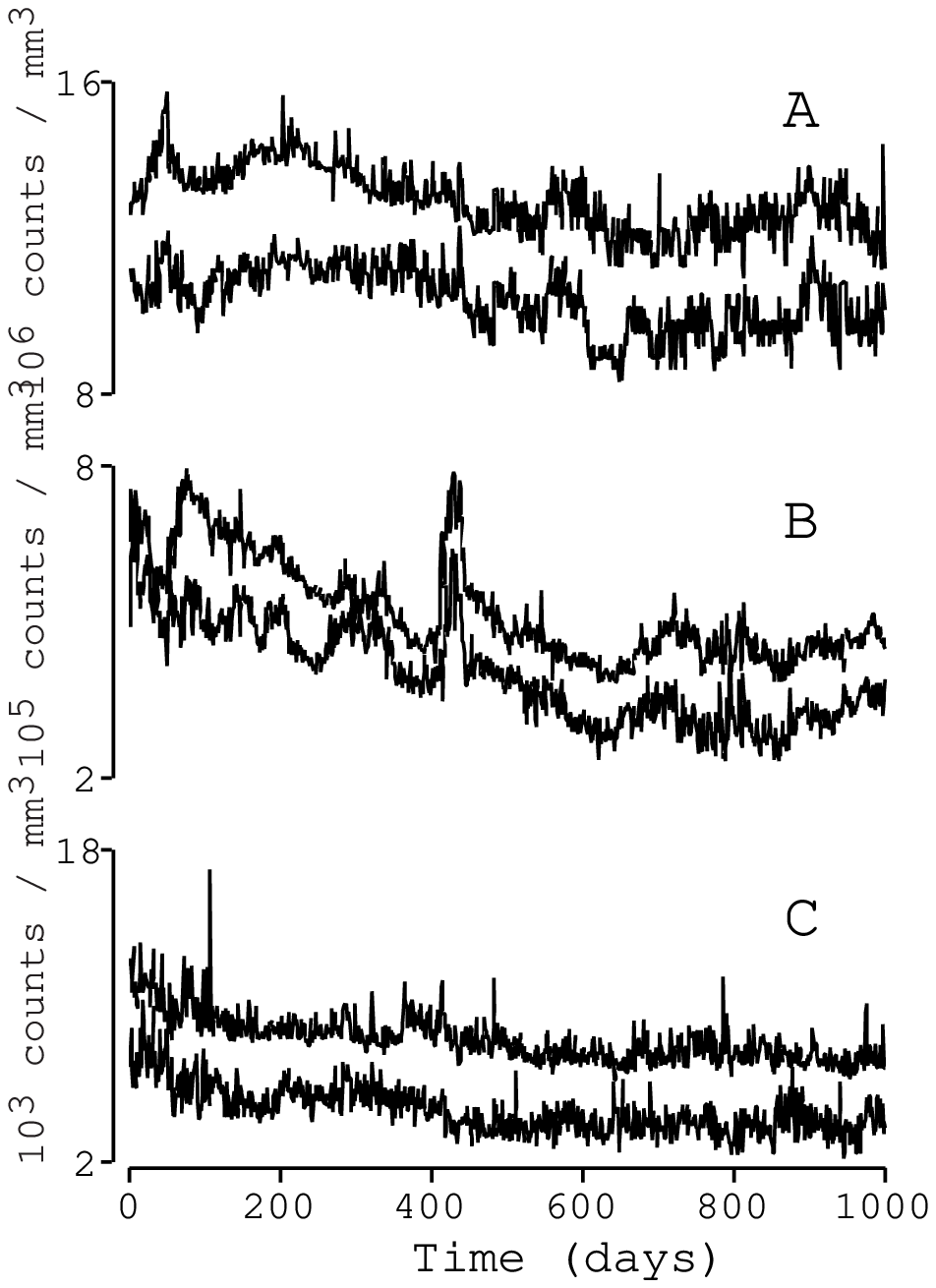,width=3.00truein}}
\caption{\footnotesize{ Daily blood cell counts time series recorded over one thousand days in two healthy 
sheep (identified as sheep 77 and 78). The erythrocyte (panel a), platelet (panel b), and leukocyte 
(panel c) time series of both sheep are depicted. For clarity the series corresponding to sheep 77, have 
been shifted upward by 15\% (red), 50\% (platelet) and 100\% (white) of their mean values.
Data was collected from two castrated male sheep, one year old at the beginning of data collection, fed ad-
libitum and housed in individual pens, diagnosed as healthy during the period by our in-house 
veterinarians. A few ml of blood were extracted every morning from the jugular vein over 1000 
consecutive days and red cells, platelets and white blood cells were counted using a manual (non-
automatic) procedure. 
}}\end{figure}

\begin{figure}[htbp]
\centerline{\psfig{figure=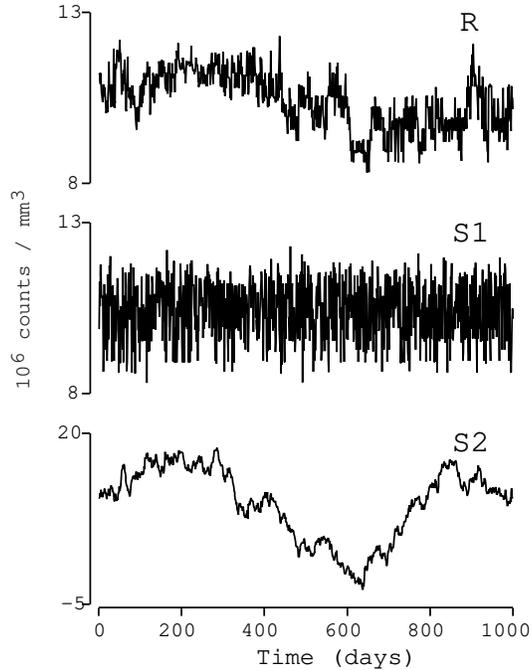,width=2.75truein}}
\caption{\footnotesize{  
Raw and surrogate series (from animal 78). The raw time series of red blood cell counts 
is depicted in the upper panel. Bottom panels denoted S1 and S2 correspond to two examples of 
potential processes (i.e., the null hypotheses) which could give rise to the observed fluctuations. The 
surrogate type 1 time series is built by randomly re-ordering the raw data, thus assuming a mean value 
contaminated with white noise having a similar density to the raw data. Surrogate type 2 preserves the 
day-to-day increments, and is built by re-ordering and integrating the day-to-day red cell count 
differences, therefore resulting in a random walk-like signal. Simple inspection reveals that the 
dynamics of the raw data lies between the two surrogates, departing from the mean value significantly 
less than the S2 surrogate (which is representative of a random walk signal) but more than the S1 
signal. 
}}\end{figure}

\begin{figure}[htbp]
\vspace{-1.75truein}
\centerline{\psfig{figure=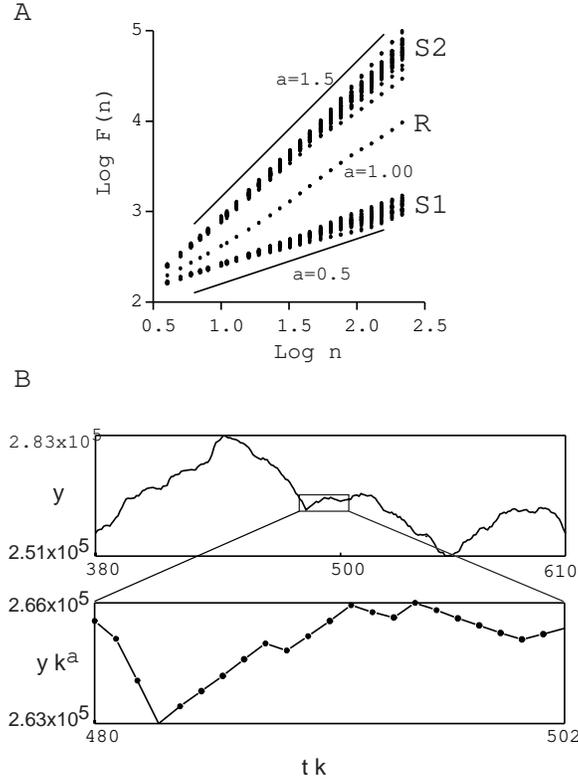,width=3.00truein}}
\caption{\footnotesize{Panel a: Log-Log plot of the mean fluctuation function F(n) (for red blood cells in 
animal 78). R, S1 and S2 denote detrended fluctuation analyses of the time series of the raw data and 
surrogates plotted in Fig. 2 . (30 random realizations for S1 and S2). The straight lines with slopes
$\alpha= 0.5$ and $\alpha= 1.5$ indicate the expected scaling for white noise and random walk. The red cell scaling 
exponent of $\alpha=1.00$ is clearly different from that of the surrogate series.
Panel b: A graphical demonstration of scale-invariance showing the self-affine transformation that 
preserves the statistical properties of the trace. Using $\alpha=1.00$ calculated for the raw data (``R'') in panel 
a, if one expands the time axis by $k$ and the $y$ axis by $k^{\alpha}$ then over both observation periods (230 and 23 
days) it should be possible to observe an excursion which occupies the entire range of $y$ as indeed is 
seen in the figure. 
}}\end{figure}

\end{document}